\documentstyle[prd,aps,multicol]{revtex}
\widetext
\renewcommand{\narrowtext}{\begin{multicols}{2} \global\columnwidth20.5pc}
\renewcommand{\widetext}{\end{multicols} \global\columnwidth42.5pc}
\def\Rrule{\vspace{-0.1in}\hfill\vrule depth1em height0pt \vrule
  width3.5in height.2pt depth.2pt\vspace*{-0.1in}}
\def\bml{\begin{mathletters}}
\def\eml{\end{mathletters}}
\def\beq{\begin{equation}}
\def\eeq{\end{equation}}
\def\bea{\begin{eqnarray}}
\def\eea{\end{eqnarray}}
\def\ba{\begin{array}}
\def\ea{\end{array}}
\def\noi{\noindent}
\def\to{\rightarrow}
\def\a{\alpha}

\def\l{\lambda}
\def\m{\mu}
\def\n{\nu}
\def\z{\zeta}
\def\f{\tilde{f}}
\def\r{\tilde{r}}
\def\q{\tilde{q}}
\def\C{\tilde{c}}
\def\e{{\rm e}}
\def\tr{\,{\rm tr}\,}
\def\diag{\,{\rm diag}\,}
\def\re{\,{\rm Re}\,}
\def\pf{\,{\rm Pf}\,}
\begin{document}
\preprint{
TIT-HEP-440}
\draft
\title{Massive chiral random matrix ensembles
at ${\bf \beta}$ = 1 \& 4 :\\
Finite-volume QCD partition functions}
\author{Taro Nagao}
\address{
Department of Physics, Graduate School of Science,
Osaka University, Toyonaka, Osaka 560-0043, Japan
}
\author{Shinsuke M. Nishigaki}
\address{
Department of Physics, Faculty of Science,
Tokyo Institute of Technology,
Oh-okayama, Meguro, Tokyo 152-8551, Japan
}
\date{January 21, 2000}
\maketitle
\begin{abstract} 
In a deep-infrared (ergodic) regime,
QCD coupled to massive pseudoreal and real quarks
are described by chiral orthogonal and symplectic ensembles
of random matrices.
Using this correspondence,
general expressions for the QCD partition functions
are derived in terms of microscopically rescaled mass variables.
In limited cases, 
correlation functions of Dirac eigenvalues and
distributions of the smallest Dirac eigenvalue are
given as ratios of these partition functions.
When all masses are degenerate,
our results
reproduce the known expressions for
the partition functions of zero-dimensional $\sigma$ models.
\end{abstract}

\renewcommand{\thefootnote}{\fnsymbol{footnote}}
\setcounter{footnote}{0}
\narrowtext
The low energy dynamics of quantum chromodynamics
is dictated by confinement of colored particles
and spontaneous breakdown of the chiral symmetry
\cite{Wei}.
While the vector part of the flavor symmetry
in a vectorial theory such as QCD
is protected by Vafa-Witten theorem \cite{VW},
its axial part is considered to be maximally broken
by the quark condensate.
Accordingly the pattern of global symmetry breaking
of QCD coupled to $n(\geq 2)$ massless quarks
falls onto either of the following three classes
\cite{LS,SmV}:
\bml
\bea
N_c\geq 3,\ \mbox{fund.}\
&:&\ SU(n)\times SU(n) \to SU(n),\\
N_c=2,\ \mbox{fund.}\
&:&\ SU(2n) \to Sp(2n),\\
N_c\geq 2,\ \mbox{adj. }\,\
&:&\ SU(n) \to SO(n),
\eea
\label{chiSB}
\eml
where `fund.' and `adj.' stand for the representations
of the $SU(N_{c})$ gauge group to which quark fields belong.
These three classes are assigned
Dyson indices $\beta=2, 1, 4$, respectively,
according to the anti-unitary symmetry of associated
Euclidean Dirac operators \cite{V3foldway}.
In the vicinity of the chiral limit where
the quark masses $m$ are much smaller than the
typical hadronic scale $\Lambda_{\rm QCD}$,
the theory is effectively described by
a non-linear $\sigma$ model over the coset manifold,
without involving gluons and quarks that are confined.
The effective Lagrangian consists of terms made of
the Goldstone pion field $U(x)$ and the quark mass matrix,
consistent with the flavor symmetry (\ref{chiSB}) \cite{GL85}.
Furthermore, in the `ergodic regime'
where the linear dimension $L$ of the system
is much smaller than the pion Compton length
$1/m_{\pi}\sim 1/\sqrt{m\Lambda_{\rm QCD}}$,
only the zero mode of $U(x)$ dominates, so that the effective
`finite-volume' partition function
in the presence of the $\theta$ angle
simplifies into a finite-dimensional integral:
\cite{LS,SmV}
\bml
\begin{eqnarray}
&&Z^{(2)}(\theta; M) =
\int_{SU(n)} \!\!\!\!\!\!\!\!\!dU\,\exp
\left(\re \tr \e^{i\theta/n} M U^\dagger \right),\\
&&Z^{(1)}(\theta; M) =
\int_{SU(2n)} \!\!\!\!\!\!\!\!\!\!\! dU\,
\exp\biggl(
\frac12 \re \tr\e^{i\theta/n}  M U {\bf J} U^T \biggr),\\
&&Z^{(4)}(\theta; M) =
\int_{SU(n)}  \!\!\!\!\!\!\!\!\! dU\,
\exp\left( \re  \tr\e^{i\theta/(N_{c}n)} M U U^T \right).
\end{eqnarray}
\label{fvpftheta}
\eml
Here the rescaled mass matrices $M$ are
\begin{eqnarray}
M = &&\diag(\m_1, \ldots, \m_n)\ \ \ \ \ \ \ \ \ \ \ \ \,
 (\beta=2,4),
\nonumber\\
M = &&\diag(\m_1, \ldots, \m_n)\otimes J\ \ \ \ \ \ \
 (\beta=1),
\\
&&{\bf J}=\openone_n \otimes J,\
J=\left( 
\ba{cc}
 0 & 1 \\
-1 & 0
\ea
\right), 
\nonumber\\
L^4\to&&\infty,\ \ m_i\to 0,\ \ \mu_{i}= \Sigma L^4 m_{i} : \mbox{fixed},
\label{mu}
\end{eqnarray}
and $\Sigma$ stands for the quark condensate in the chiral limit.
The integrals are extended from cosets to full unitary groups
(for $\beta=1, 4$),
and their Haar measures $dU$ are normalized such that
$\int dU=1$.
After Fourier transformation, the partition function
in a sector with a topological charge $\nu$
takes the form:
\bml
\begin{eqnarray}
&&Z^{(2)}_\nu(M) =
\int_{U(n)} \!\!\!\!\!\!\!dU\,\det\nolimits^{-\nu} U\, \exp\left(
\re \tr M U^\dagger \right),\\
&&Z^{(1)}_\nu(M) =\int_{U(2n)}
\!\!\!\!\!\!\!\!\! dU\,\det\nolimits^{-\nu} U\,
\exp\biggl(
\frac12 \re \tr M U {\bf J} U^T \biggr),\\
&&Z^{(4)}_\nu(M) =
\int_{U(n)}  \!\!\!\!\!\!\!
dU\,\det\nolimits^{-2\nu} U\, \exp\left(
\re \tr M U U^T \right).
\end{eqnarray}
\label{fvpf}
\eml
\noindent
For the $\beta=4$ case, $\nu$ is understood to be substituted by $\nu N_c$
\cite{SmV}.
With a help of Itzykson-Zuber-like formula,
an explicit form of the integral, which can be considered as
a `matrix Bessel function', was obtained for $\beta=2$
\cite{GW,JSV}:
\bea
Z^{(2)}_\nu(\m_1,\ldots,\m_n)&=& 2^{n(n-1)\over 2}\!
\prod_{k=0}^{n-1} k!
\frac{
\det\limits_{1\leq i,j\leq n} \m_i^{j-1} I_{\nu+j-1}(\m_i)}{
\Delta(\m_1^2,\ldots,\m_n^2)},
\nonumber\\
\Delta(x_1,\ldots,x_{n})&\equiv& \prod_{i > j}^n(x_{i}-x_{j}),
\label{Z2}
\eea
and namely for identical  $\m_i$'s \cite{LS},
\beq
Z^{(2)}_\nu(\underbrace{\m,\ldots,\m}_n)
= \det\limits_{-\frac{n-1}{2} \leq i,j \leq \frac{n-1}{2}}
I_{\nu+j-i}(\m).
\eeq
On the other hand, due to technical difficulties,
only the case with completely
degenerated masses was worked out for $\beta=1$ and 4 \cite{SmV}:
\beq
Z^{(1)}_\nu(\underbrace{\m,\ldots,\m}_n)
=
\frac{1}{(2n-1)!!}\,
{}_{{}_{-{\scriptstyle n}+\frac{1}{2} \leq {\scriptstyle  i,j}
\leq {\scriptstyle n} - \frac{1}{2}}}
\!\!\!\!\!\!\!\!\!\!\!\!\!\!\!\!\!\!\!\!\! \pf\ \ \ \ \ \,\,
(j-i) I_{\nu+j+i}(\m),
\label{Z1sigma}
\eeq
\bml
\bea
Z^{(4)}_\nu(\underbrace{\m,\ldots,\m}_n)
&=&(n-1)!! \pf (A) \ \ \ \ \ \ \ \ (n:\ \mbox{even})
\label{Z4evensigma}\\
&=&
n!!\pf \left(
\begin{array}{cc}
A &  b \\
-b^T & 0
\end{array}
\right)
\ \ \ \ (n:\ \mbox{odd}),
\\
A^{ij}&\equiv&
\sum_{k=-\infty}^{\infty}\frac{1}{2k+1}
I_{\nu+i+k+\frac12}(\m)I_{\nu+j-k-\frac12}(\m),\nonumber\\
b^i&\equiv& I_{\nu+i}(\m),
\ \ 
-\frac{n-1}{2} \leq i,j\leq \frac{n-1}{2}.
\nonumber
\eea
\label{Z4sigma}
\eml
\ \\
Although the above expressions are primarily valid for $n\geq 2$,
they can be extended to $n=1$,
as a separate consideration for this mass-gapped case
leads to \cite{LS}:
\beq
Z^{(\beta)}_\nu(\mu)=I_\nu(\mu),
\eeq
irrespective of the values of $\beta$.
Once the partition functions are computed in a closed form
as the above,
physical quantities such as the topological susceptibility
\beq
\langle \nu^2 \rangle =
\frac{\sum_{\nu} \nu^2 Z^{(\beta)}_\nu(\{\mu\})}{
\sum_{\nu} Z^{(\beta)}_\nu(\{\mu\})}
\eeq
are expressed in terms of them.

An important observation made by Verbaarschot and collaborators
\cite{ShV,HV95}
is that the finite-volume partition functions (\ref{fvpf}) can as well be
derived from models much simpler than QCD,
Chiral Random Matrix Ensembles ($\chi$RMEs,
for a recent review see Ref.\cite{Vreview}):
\beq
{\cal Z}^{(\beta)}_\nu (\{m\})=
\int dW \e^{-\beta \tr V(W^\dagger W)}
\prod_{i=1}^n
\det 
\left(
\ba{cc}
m_i & W \\
-W^\dagger & m_i
\ea
\right),
\label{ZchiRME}
\eeq
where the integrals are over complex, real, and quaternion real
$N\times(N+\nu)$ matrices $W$ for $\beta=2,1,4$, respectively, and
it is understood for $\beta=4$ that twofold degenerated eigenvalues
in the determinant are only counted once.
Their proofs consist of the `color-flavor transformation'
\cite{Zir} that converts the integration variables into
$n\times n$ matrices and the saddle point method under which
\beq
N\to \infty,\ \ \ m_i \to 0, \ \ \ \mu_i\equiv\pi \rho(0) m_i
\mbox{ : fixed}.
\label{microlim}
\eeq
Here $\rho(0)$ stands for the spectral density of the random matrix
${\cal D}=\left({\ \ 0 \ \ \ \ W  \atop -W^\dagger\ \  0 }\right)$:
\beq
\rho(\lambda)=\langle \tr \delta(\lambda-i{\cal D}) \rangle,
\eeq
at the origin.
The $\chi$RME is motivated by
the microscopic theory (Euclidean QCD) on a lattice,
with a crude simplification of replacing
matrix elements of the anti-Hermitian Dirac operator
$/\!\!\!\!D= (\partial_\mu + i A_\mu)\gamma_\mu$
by random numbers ${\cal D}$ generated according to
the weight $\e^{-\beta \tr V(W^\dagger W)}$.
Under this correspondence, the microscopic limit (\ref{microlim})
is equivalent to Leutwyler-Smilga limit (\ref{mu}),
since the size $N$ of the matrix $W$ is interpreted as
the number of cites $L^4$ of the lattice on which QCD is discretized,
and the Dirac spectral density at zero virtuality $\rho(0)$
is related to the quark condensate by Banks-Casher relation
$\Sigma=\pi\rho(0)/L^4$ \cite{BC}.
These $\chi$RMEs are technically more suited for the computation
of microscopic spectral correlations of Dirac operators
$/\!\!\!\!D \sim {\cal D}$, than the zero-dimensional $\sigma$ models
(\ref{fvpf}) where 
inevitable introduction of a probe quark pair (partial quenching)
\cite{DOTV,TV} brings forth additional complication.
In the chiral limit $\mu\equiv 0$,
the eigenvalue correlation functions \cite{NSl,AST,V_chGOE,For,NF95}
as well as the smallest eigenvalue distributions \cite{FH,SlN,TW,NF98}
have been computed for $\chi$RMEs with all three values of $\beta$.
On the other hand, in the presence of finite and generic $\mu$'s,
these quantities have so far been analytically computed
only for the $\beta=2$ case \cite{DN98a,WGW,JNZ,NDW}
(For $\beta=4$, the smallest eigenvalue distribution in the presence
of $\mu$'s was numerically computed in Ref.\cite{BBMW},
and the partition function with four degenerate $\mu$'s was analytically
treated in Ref.\cite{AD99}).
We shall treat the remaining cases, chiral orthogonal ($\beta=1$) and
symplectic ($\beta=4$) ensembles with finite mass parameters.
This Letter is devoted to the computation of
partition functions of these ensembles.
As corollaries, we provide closed expressions for
the smallest eigenvalue distribution for $\beta=1$ and odd $\nu$, and
the microscopic eigenvalue correlation functions for $\beta=4$.
The microscopic eigenvalue correlation functions for the $\beta=1$ case
will be presented in a separate publication \cite{NN}.

We start by expressing the partition function (\ref{ZchiRME})
of the $\chi$RME in terms of eigenvalues $x_i=\lambda_i^2$ of the positive
definite
matrix $W^\dagger W$ (up to a constant independent of $m$ and $\nu$):
\begin{eqnarray}
{\cal Z}^{(\beta)}_\nu (\{ m \})&=&
\biggl(\prod_{i=1}^n m_i^\nu \biggr) \Xi^{(\beta)}_\nu (\{ m \}),
\label{ZchiRME_x}\\
\Xi^{(\beta)}_\nu (\{ m \})&=&
\int_0^\infty \prod_{j=1}^N \left( dx_j \, w(x_j; \{m\}) \right)
|\Delta(x_1,\ldots,x_N)|^\beta , \nonumber\\
w(x; \{m\})&=& {\rm e}^{-\beta V(x)} x^{{\beta}(\nu+1)/2 -1} \prod_{i=1}^n
(x+m_i^2) .
\label{Xi}
\end{eqnarray}
Since the partition function is even under $\nu\to-\nu$,
we have set $\nu$ non-negative integer, without loss of generality.
We note that all spectral correlation functions of the
orthogonal and symplectic ensembles can be constructed from
the scalar kernel of the unitary ensemble sharing the
same weight function $w(x)$ \cite{SV,Wid}.
Since the scalar kernel
in the microscopic limit (\ref{microlim})
is known to be insensitive to the details of the potential $V(x)$
either in the absence \cite{ADMN,KF} or
in the presence of finite $\mu$'s \cite{DN98a},
it suffices to concentrate onto Laguerre (chiral Gaussian)
ensembles, $V(x)= x$.
This choice leads to
\beq
\rho(\lambda)=\frac{2}{\pi}\sqrt{2N-\lambda^2}.
\eeq

We sketch the skew-orthogonal polynomial method \cite{Meh}
as employed in Ref.\cite{NF98}, to which we leave details.
As the final result in the microscopic limit is insensitive to the
parity of $N$, we consider only even $N$ henceforth.

\subsection{orthogonal ensemble}
We use the identity
\bea
&&\Delta(x_1,\ldots,x_N)
\prod_{j=1}^N \prod_{i=1}^n (x_j-x_{N+i}) \nonumber\\
&=&
\frac{\Delta(x_1,\ldots,x_{N+n})}{\Delta(x_{N+1},\ldots,x_{N+n})}
=
\frac{\det\limits_{1\leq i,j \leq N+n} R_{i-1}(x_j)}%
{\Delta(x_{N+1},\ldots,x_{N+n})},
\label{detoverdet1}
\eea
where $R_{i}(x)$ is an arbitrary monic polynomial of the $i$-th order,
and $x_{N+i}\equiv -m_i^2 \leq 0$.
We take $\{R_{i}(x)\}$ to be skew-orthogonal
\beq
\langle R_{2i},R_{2j+1} \rangle_R= -\langle R_{2j+1},R_{2i} \rangle_R
=h_i \delta_{ij},\ \
\mbox{others}=0,
\label{skew2}
\eeq
with respect to the antisymmetric product
\bea
\langle f,g \rangle_R &=&
\int_0^\infty dx\, x^{\frac{\nu-1}{2}} \e^{-x} g(x)
\int_0^x dy\, y^{\frac{\nu-1}{2}} \e^{-y} f(y) \nonumber\\
&&- (f\leftrightarrow g).
\eea
When eq.(\ref{detoverdet1}) is integrated
over $x_1,\ldots,x_N$ with the weight
$\prod_{i=1}^N (\e^{-x_i} x_i^{\frac{\nu-1}{2}})$
in a cell $0\leq x_1\leq x_2 \leq \cdots \leq x_N$,
it can be neatly expressed as a Pfaffian,
due to the skew orthogonality (\ref{skew2}) \cite{NF98}:
\bml
\bea
&&\Xi^{(1)}_\nu(\{m\})
=
\frac{ \left(\prod_{i=0}^{\frac{N+n}{2}-1} h_{i}\right)
\pf (F)}{\Delta(m_1^2,\ldots,m_n^2)}
\ \ \ \ \ \ \ \ (n:\mbox{even}) \\
&&\ \ \ \ \ = 
\frac{\left(\prod_{i=0}^{[\frac{N+n}{2}]-1} h_{i}\right)
\pf 
\left(
\ba{cc}
 F & R\\
-R^T & 0
\ea
\right)
}{\Delta(m_1^2,\ldots,m_n^2)}
\ \ \ (n:\mbox{odd}),
\eea
\label{Xi1}
\eml 
where
\begin{eqnarray*}
&&F^{ij}=
\sum_{k=0}^{[\frac{N+n}{2}]-1}
\frac{R_{2k}(-m^2_i)R_{2k+1}(-m^2_j)-
(i\leftrightarrow j)
}{h_{k}}, \\
&&
R^i=R_{N+n-1}(-m^2_i),\ \ \ \ \ \ \ \ \ \ 1\leq i,j\leq n.
\end{eqnarray*}
Explicit forms of the monic skew-orthogonal polynomials
and their norms
are known as \cite{NW}:
\bea
&&R_{2k}(x)=
-\frac{(2k)!}{2^{2k+1}}\frac{d}{dx} L^{(\nu-1)}_{2k+1}(2x),
\nonumber\\
&&R_{2k+1}(x)=
-\frac{(2k+1)!}{2^{2k+1}}L^{(\nu-1)}_{2k+1}(2x)\\
&&\ \ \ \ \ \ \ \ \ \ \ \
-\frac{(2k)!}{2^{2k+2}}(2k+\nu)\frac{d}{dx} L^{(\nu-1)}_{2k}(2x),
\nonumber\\
&&h_k=2^{-4k-\nu}(2k)!(2k+\nu)!.
\nonumber
\eea
In the microscopic limit (\ref{microlim})
with $\mu_i=2\sqrt{2N} m_i$ fixed,
the sum over the indices $k$
becomes an integral, and Laguerre polynomials
approach modified Bessel functions:
\beq
L^{(\a)}_k(x) \sim \left(\frac{k}{-x}\right)^{\frac{\a}{2}}
I_\a(2\sqrt{-kx})  \ \ \Bigl(x=O(\frac{1}{k})<0 \Bigr).
\eeq
Then the partition function is expressed as
\bml
\bea
Z^{(1)}_\nu(\{\m\})&=& \biggl( \prod_{i=1}^n \mu_i^\nu \biggr)
\xi^{(1)}_\nu(\{\m\}),
\nonumber\\
\xi^{(1)}_\nu(\{\m\})&=&
c_n
\frac{\pf (f)}{\Delta(\m_1^2,\ldots,\m_n^2)}
\ \ \ \ \ \ \ \ \ \, (n:\mbox{even}) \\
&=& 
c_n
\frac{\pf 
\left(
\ba{cc}
 f   & r\\
-r^T & 0
\ea
\right)
}{\Delta(\m_1^2,\ldots,\m_n^2)}
\ \ \ \ \ \ \ \ (n:\mbox{odd}),
\eea
\label{Z1}
\eml
where
\begin{eqnarray*}
c_n
&=& 
(-1)^{{n(n-1)\over 2}}
2^{n^2-1}
(n-1)! \prod_{k=0}^{n-2} (2k+1)!
\ \ \ \ (n:\mbox{even}) \\
&=&
(-1)^{{n(n-1)\over 2}}
2^{{(n-1)(7n+11)\over 8}}
(n-1)!! \prod_{k=0}^{n-2} (2k+1)!
\nonumber\\
&&
\ \ \ \ \ \ \ \ \ \ \ \ \ \ \ \ \ \
\ \ \ \ \ \ \ \ \ \ \ \ \ \ \ \ \ \
\ \ \ \ \ \ \ \ \ \ \ \ \ \ \ \  (n:\mbox{odd}) ,\\
f^{ij}
&=&\int_0^1 dt\, {t}^2
\frac{I_{\nu-1}({t}\mu_i)}{\mu_i^{\nu-1}}
\frac{I_{\nu}({t}\mu_j)}{\mu_j^{\nu}}
-(i \leftrightarrow j), \\
r^i&=&\frac{I_{\nu}(\mu_i)}{\mu_i^{\nu}},
\ \ \ \ \ \ \ \ \ \ 1\leq i,j\leq n.
\end{eqnarray*}
The constant $c_n$ is conveniently determined as the above
by requiring the small-$\mu$ behavior be in accord with
that of the zero-dimensional $\sigma$ model,
\beq
Z^{(1)}_\nu(\underbrace{\m,\ldots,\m}_n)
\simeq
\left(\frac{\m}{2} \right)^{n\nu}
\prod_{k=0}^{n-1} \frac{(2k)!}{(2k+\n)!}
\ \ \ \; (\m\ll 1).
\eeq

\subsection{symplectic ensemble}
We concentrate on the case with an even $n(\equiv 2a)$
number of flavors and
pairwise degenerated mass parameters,
corresponding to adjoint Dirac fermions in the QCD context.

We use the identity
\bea
&&\Delta(x_1,\ldots,x_N)^4
\prod_{j=1}^N \prod_{i=1}^a (x_j-x_{N+i})^2 \nonumber\\
&&=
\frac{
\left|
\ba{rrrrrc}
1 & x_1 & x_1^2  &\cdots & &x_1^{2N+a-1} \\
0 & 1 & 2x_1 & \cdots & (2N+a-1)\!\!\!& x_1^{2N+a-2} \\
\vdots &  \vdots &  \vdots & \ddots & & \vdots \\
\vdots &  \vdots &  \vdots & & \ddots & \vdots \\
1 & x_N & x_N^2 & \cdots & & x_N^{2N+a-1} \\
0 & 1 & 2x_N & \cdots & (2N+a-1)\!\!\!& x_N^{2N+a-2} \\
1 & x_{N+1} & x_{N+1}^2 &\cdots & &x_{N+1}^{2N+a-1} \\
\vdots &  \vdots & \vdots & \ddots&  & \vdots \\
1 & x_{N+a} & x_{N+a}^2  & \cdots&  & x_{N+a}^{2N+a-1}
\ea
\right|
}{\Delta(x_{N+1},\ldots,x_{N+a})}\nonumber\\
&&=
\frac{
\left|
\ba{llll}
Q_0(x_1) & Q_1(x_1)  & \cdots & Q_{2N+a-1}(x_1) \\
Q'_0(x_1) & Q'_1(x_1)  & \cdots & Q'_{2N+a-1}(x_1) \\
\vdots & \vdots &  \ddots  & \vdots \\
\vdots & \vdots &  \  & \vdots \\
Q_0(x_N) & Q_1(x_N) & \cdots & Q_{2N+a-1}(x_N) \\
Q'_0(x_N) & Q'_1(x_N)  & \cdots & Q'_{2N+a-1}(x_N) \\
Q_0(x_{N+1}) & Q_1(x_{N+1}) & \cdots & Q_{2N+a-1}(x_{N+1}) \\
\vdots & \vdots &  \ddots & \vdots \\
Q_0(x_{N+a}) & Q_1(x_{N+a}) & \cdots & Q_{2N+a-1}(x_{N+a})
\ea
\right|
}{\Delta(x_{N+1},\ldots,x_{N+a})},
\label{detoverdet4}
\eea
where $Q_{i}(x)$ is an arbitrary monic polynomial of the $i$-th order,
and $x_{N+i}\equiv -m_i^2 \leq 0$.
We take $\{Q_{i}(x)\}$ to be skew-orthogonal
\beq
\langle Q_{2i},Q_{2j+1} \rangle_Q= -\langle Q_{2j+1},Q_{2i} \rangle_Q
=h_i \delta_{ij},\ \
\mbox{others}=0,
\label{skew1}
\eeq
with respect to the antisymmetric product
\bea
\langle f,g \rangle_Q &=&
\int_0^\infty dx\, x^{2\nu+1} \e^{-4x} (f(x)g'(x) -f'(x)g(x)).
\eea
When eq.(\ref{detoverdet4}) is integrated
over $x_1,\ldots,x_N$ with the weight
$\prod_{i=1}^N (\e^{-4x_i} x_i^{2\nu+1})$,
it can be neatly expressed as a Pfaffian,
due to the skew orthogonality (\ref{skew1}) \cite{NF98}:
\bml
\bea
&&\Xi^{(4)}_\nu(\{m\})
=
\frac{ \left(\prod_{i=0}^{N+\frac{a}{2}-1} h_{i}\right)
\pf (F)}{\Delta(m_1^2,\ldots,m_{a}^2)}
\ \ \ \ \ \ (a:\mbox{even}) \\
&&\ \ \ = 
\frac{\left(\prod_{i=1}^{N+[\frac{a}{2}]-1} h_{i}\right)
\pf 
\left(
\ba{cc}
F & Q\\
-Q^T & 0
\ea
\right)
}{\Delta(m_1^2,\ldots,m_{a}^2)}
\ \ \ (a:\mbox{odd}),
\eea
\label{Xi4}
\eml 
where
\begin{eqnarray*}
&&F^{ij}=
\sum_{k=0}^{N+[\frac{a}{2}]-1}
\frac{Q_{2k}(-m^2_i)Q_{2k+1}(-m^2_j)-
(i\leftrightarrow j)}{h_{k}}, \\
&&
Q^i=Q_{2N+a-1}(-m^2_i),
\ \ \ \ \ \ 1\leq i,j \leq a.
\end{eqnarray*}
Explicit forms of the monic skew-orthogonal polynomials
and their norms
are known as \cite{NW}:
\bea
&&Q_{2k}(x)=
\frac{k!(k+\nu)!}{2^{2k}}
\sum_{l=0}^k \frac{(2l-1)!!}{2^{2l}(l+\nu)!}L^{(2\nu)}_{2l}(4x),
\nonumber\\
&&Q_{2k+1}(x)=
-\frac{(2k+1)!}{2^{4k+2}}L^{(2\nu)}_{2k+1}(4x),\\
&&h_k=2^{-8k-4\nu-4}(2k+1)!(2k+2\nu+1)!\nonumber.
\eea
In the microscopic limit
with $\m_i=2\sqrt{2N} m_i$ fixed,
the sums over the indices $k$ and $l$
become integrals, and Laguerre polynomials
approach modified Bessel functions.
Then the partition function is expressed as
\bml
\bea
Z^{(4)}_\nu(\{\m\})&=& \biggl( \prod_{i=1}^a \mu_i^{2\nu} \biggr)
\xi^{(4)}_\nu(\{\m\}),
\nonumber\\
\xi^{(4)}_\nu(\{\m\})&=&
c_a
\frac{\pf (f)}{\Delta(\m_1^2,\ldots,\m_{a}^2)}
\ \ \ \ \ \ \; (a:\mbox{even}) \\
&=&
c_a
\frac{\pf 
\left(
\ba{cc}
f & q\\
-q^T & 0
\ea
\right)
}{\Delta(\m_1^2,\ldots,\m_{a}^2)}
\ \ \ \ \ (a:\mbox{odd}),
\eea
\label{Z4}
\eml
where
\begin{eqnarray*}
c_a
&=& 
(-1)^{{a(a-1)\over 2}}
\prod_{k=0}^{a-1} (2k+1)! \\
f^{ij}
&=&
\int_0^1 dt\,t \frac{I_{2\nu}(2{t}\mu_i)}{\mu_i^{2\nu}}
\int_0^1 du \frac{I_{2\nu}(2{t}u\mu_j)}{\mu_j^{2\nu}}
-(i \leftrightarrow j), \\
q^i&=&\int_0^1 dt \frac{I_{2\nu}(2t\mu_i)}{\mu_i^{2\nu}},
\ \ \ \ \ \ \ \ \ 1\leq i,j\leq a.
\end{eqnarray*}
The constant $c_a$ is conveniently determined as the above
by requiring the small-$\mu$ behavior be in accord with
that of the zero-dimensional $\sigma$ model,
\beq
Z^{(4)}_\nu(\underbrace{\m,\ldots,\m}_{2a} )
\simeq
\m^{2a\nu}
\prod_{k=0}^{a-1} \frac{(2k+1)!}{(2k+2\n+1)!}
\ \ \  (\m\ll 1).
\eeq

It remains to confirm whether the above expressions for
the $\chi$RME partition functions agree with
those of the zero-dimensional $\sigma$ models.
By taking all $\m$'s to be identical, we obtain
for $\beta=1$:
\bml
\begin{eqnarray}
Z^{(1)}_\nu(\underbrace{\m,\ldots,\m}_n)
&&= 
\C_n
\frac{\pf (\f)}{\m^{(n/2-1)n}}
\ \ \ \ \ \ \ \ \ \ \ \ (n:\mbox{even}) \\
&&= 
\C_n
\frac{\pf\left(
\ba{cc}
 \f   & \r\\
-\r^T & 0
\ea
\right)}{\m^{(n-1)^2/2}} \ \ \ \ \!(n:\mbox{odd}),
\end{eqnarray}
\eml
where
\begin{eqnarray*}
\C_n
&=&
(-1)^{{n(n-1)\over 2}}
2^{
(n-1)(\frac{n}{2}+1)
}\prod_{k=0}^{\frac{n}{2}-1}
\frac{(2k+n-1)!}{(2k)!}\ (n:\mbox{even}) \\
&=&
(-1)^{{n(n-1)\over 2}}
2^{
{(n-1)(3n+11)\over 8}
}
\frac{\prod_{k=0}^{\frac{n-3}{2}}(2k+n)!}{\prod_{k=0}^{\frac{n-5}{2}}(2k+1)!
}
\ \ \ \ (n:\mbox{odd}) ,\\
\f^{ij}&=&\int_0^1 dt\,t^{{i+j+2}}
I_{\nu+i-1}(t\m)I_{\nu+j}(t\m)-
(i\leftrightarrow j),\\
\r^{i}&=&I_{\nu+i}(\m), \ \ \ \ \ \ \ \ 0\leq i,j \leq n-1 ,
\end{eqnarray*}
and for $\beta=4$:
\bml
\begin{eqnarray}
Z^{(4)}_\nu(\underbrace{\m,\ldots,\m}_{2a})&=&
\C_a
\frac{\pf (\f)}{\m^{(a-1)a/2}}
\ \ \ \ \ \ \ \ \ \ \; (a:\mbox{even}) \\
&=&
\C_a 
\frac{
\pf  \left(
\ba{cc}
\f & \q\\
-\q^T & 0
\ea
\right)}{\m^{(a-1)a/2}}
\ \ \ (a:\mbox{odd}),
\end{eqnarray}
\eml
where
\begin{eqnarray*}
\C_a
&=&
(-1)^{{a(a-1)\over 2}}
\prod_{k=0}^{a-1} \frac{(2k+1)!}{k!} ,\\
\f^{ij}&=&
\int_0^1 dt\,t^{{i+j+1}}I_{2\nu+i}(2{t}\m)
\int_0^1 du\,u^j I_{2\nu+j}(2{t}u\m) \\
&&- 
(i\leftrightarrow j), \nonumber\\
\q^{i}&=&
 \int_0^1 dt\,t^i I_{2\nu+i}(2t\m) ,
\ \ \ \ 0\leq i,j \leq a-1 .
\end{eqnarray*}
We have numerically checked that, despite the appearances,
the above expressions are identical to
eqs.(\ref{Z1sigma}) and (\ref{Z4evensigma}).
Together with the $\beta=2$ case
that has previously been confirmed \cite{Dam,NDW},
they explicitly show the equivalence between
the $\chi$RMEs and the $\sigma$ models
in Leutwyler-Smilga limit.

\subsection{smallest eigenvalue distribution}
The probability of finding no eigenvalue
in the interval $0\leq x < s$ is given by
\beq
{E}^{(\beta)}_\nu(s;\{ m \})=
\frac{
\int_s^\infty \prod_{j=1}^N \left( dx_j \, w(x_j; \{m\}) \right)
|\Delta(\{x\})|^\beta}{
\int_0^\infty \prod_{j=1}^N \left( dx_j \, w(x_j; \{m\}) \right)
|\Delta(\{x\})|^\beta}.
\eeq
The integral domain in the numerator can be traded to
$[0,\infty)$ with the weight function shifted by $s$,
$w(x+s; \{m\})$.
If the exponent in the weight function
$\frac{\beta}{2}(\nu+1) -1$ is an integer
(excluding the case with $\beta=1$ and $\nu$ even),
we can utilize the `flavor-topology duality' \cite{V98}
\beq
\Xi^{(\beta)}_\nu (m_1,\ldots,m_n)=
\Xi^{(\beta)}_{\frac{2}{\beta}-1}
(m_1,\ldots,m_n,\underbrace{0,\ldots,0}_{\frac{\beta}{2}(\nu+1)-1}),
\eeq
to express ${E}^{(\beta)}_\nu(s;\{ m \})$
in terms of the partition functions:
\bea
&&{E}^{(\beta)}_\nu(s;\{ m \})=\e^{-N\beta s}\times \nonumber \\
&&
\frac{
\Xi^{(\beta)}_{\frac{2}{\beta}-1}
(\sqrt{s+m_1^2},\ldots,\sqrt{s+m_n^2},
\overbrace{\sqrt{s},\ldots,\sqrt{s}}^{\frac{\beta}{2}(\nu+1)-1})}{
\Xi^{(\beta)}_\nu (m_1,\ldots,m_n)}.
\eea
Now we change the picture back from Laguerre to chiral Gaussian, and
take the microscopic limit with
$\zeta=\pi \rho(0) \sqrt{s}=2\sqrt{2 N s}$ and
$\mu_i=2\sqrt{2 N}m_i$ fixed.
The distribution of the smallest eigenvalue
of chiral random matrices is then given by
the first $\zeta$-derivative of ${E}^{(\beta)}_\nu$:
\bea
&&P^{(\beta)}_\nu(\zeta;\{ \m \})=-\frac{\partial}{\partial \z}
\biggl\{
\e^{-(\beta/8)\zeta^2
}\times \nonumber \\
&&
\frac{
\xi^{(\beta)}_{\frac{2}{\beta}-1}
(\sqrt{\z^2+\m_1^2},\ldots,\sqrt{\z^2+\m_n^2},
.\overbrace{{\z},\ldots,{\z}}^{\frac{\beta}{2}(\nu+1)-1})}{
\xi^{(\beta)}_\nu (\m_1,\ldots,\m_n)} \biggr\}.
\label{PZ}
\eea
For $\beta=1$ and $\nu$ odd, eqs.(\ref{PZ}) and (\ref{Z1})
suffice to express
the smallest eigenvalue distribution in a closed form.
This prediction should in future be put in comparison with
lattice QCD simulations with overlap dynamical quarks.

On the other hand, for $\beta=4$,
the partition function in the numerator
falls out of the range of this Letter, as
the number of additional flavors is odd.
A different formalism based on Fredholm determinant \cite{TW96}
might be needed in order to overcome this limitation.

\subsection{correlation function}
In the case of even $\beta$,
a $p$-level correlation function
\beq
\rho(\l_1,\ldots,\l_p;\{\m\})=
\Bigl\langle
\prod_{k=1}^p \tr \delta (\lambda_{k}-i{\cal D})
\Bigr\rangle
\eeq
is expressed by construction as a ratio
of partition functions with $n$ and $n+\beta p$ flavors
\cite{Dam,AD98}. After taking the microscopic limit,
In the $\beta=4$ case with $a$ pairs of doubly degenerated masses,
it reads:
\widetext
\beq
\rho^{(4)}_\nu(\zeta_1,\ldots,\zeta_p;\{ \m \})=
C_{a,\n}^{(p)}
\Delta(\z_1^2,\ldots,\z_p^2)^4
\prod_{k=1}^p\left(\zeta_k^3 \prod_{i=1}^a (\zeta_k^2+\m_i^2)^2 \right)
\frac{Z^{(4)}_\nu(\m_1,\m_1,\ldots,\m_a,\m_a,
\overbrace{i\z_1,\ldots,i\z_1}^{4},\ldots,
\overbrace{i\z_p,\ldots,i\z_p}^{4})}{
Z^{(4)}_\nu(\m_1,\m_1,\ldots,\m_a,\m_a)}.
\eeq
\Rrule
\narrowtext
\noi
As our derivation of the partition function (\ref{Z4})
is valid as well for negative values of $m_i^2$ or $\m_i^2$,
the above relationship suffices to express
any $p$-level correlation function in a closed form,
up to an overall constant independent of $\mu$'s.

SMN thanks P.H. Damgaard and E. Kanzieper
for communications on various stages.
This work was supported in part (SMN) by
JSPS Research Fellowships for Young Scientists, and
by Grant-in-Aid No.\ 411044
from the Ministry of Education, Science and Culture, Japan.

\widetext
\end{document}